\documentclass[3p,times]{elsarticle}
\pdfoutput=1
\usepackage{ecrc}

\usepackage{graphicx}
\usepackage{amsmath}
\usepackage{fixltx2e}
\usepackage{epstopdf}
\usepackage{subfig}
\usepackage{caption}
\volume{00}

\firstpage{1}

\journalname{Nuclear Physics A}

\runauth{G. Ortona for the ALICE collaboration}

\jid{nupha}

\usepackage{amssymb}

\usepackage[figuresright]{rotating}

\begin{document}

\begin{frontmatter}




\title{Open-charm meson elliptic flow measurement in Pb-Pb collisions at $\sqrt{s_{\rm NN}} = 2.76$~TeV with ALICE at the LHC}

\author{Giacomo Ortona for the ALICE collaboration}
\ead{gortona@cern.ch}
\ead[http://personalpages.to.infn.it/~ortona]{http://personalpages.to.infn.it/~ortona}
\fntext[alicecoll]{On behalf of the ALICE collaboration}

\address{INFN - Torino, Universit\`a degli Studi di Torino}

\begin{abstract}
The ALICE experiment is one of the four large experiments at the Large Hadron Collider (LHC), and it is dedicated to the study of ultra-relativistic heavy-ion collisions, with the goal of investigating the properties of the high-density state of QCD matter produced in these collisions.

The study of D meson azimuthal anisotropy and the measurement of its elliptic flow ($v_2$) can provide insight on the degree of thermalisation of charm quarks in the medium and on the charm hadronization mechanism.

We present the measurement of the D\textsuperscript{+} and D\textsuperscript{0} meson $v_2$ in Pb-Pb collisions at $\sqrt{s_{\rm NN}}=2.76$~TeV at the LHC with ALICE. We discuss the details of the analysis and we show the results obtained from data samples collected in 2011.
\end{abstract}


\end{frontmatter}


\section{Introduction}
\label{sec:intro}
At high energy densities, Quantum--Chromo--Dynamics predicts the formation of a state of matter in which partons are not confined anymore inside hadrons. This state of matter is called Quark Gluon Plasma (QGP). The ALICE experiment \cite{alice} is one of the four large experiments at the Large Hadron Collider (LHC) and its main experimental goal is the study and characterisation of the QGP created in the high energy Pb-Pb collisions at $\sqrt{s_{NN}} = 2.76$~TeV delivered by the LHC. 
In non-central Pb--Pb collisions the fireball shows azimuthal anisotropy with respect to the reaction plane (the plane defined by the beam direction and the impact parameter). Due to collective effects, this spatial anisotropy is translated into a momentum anisotropy driven by different pressure gradients in the in-plane and out-of-plane regions. This anisotropy can be quantified expanding the particles momentum azimuthal distribution (with respect to the event plane) in a Fourier series. At mid-rapidity the main coefficient is the second coefficient of the Fourier expansion $v_2$, also called \textit{elliptic flow} \cite{v2}. 
The measurement of heavy--flavour particles $v_2$ is particularly interesting as it can probe the degree of thermalisation of the medium, the interaction of heavy quarks with the QGP and the heavy flavours hadronization mechanism \cite{rhicflowRapp,Molnar:2004ph}. The charm $v_2$ has been measured at RHIC by the PHENIX collaboration via non-photonic electrons \cite{phenixcraa}. In ALICE, Vertexing and tracking of charged particle at central rapidity ($|\eta|<0.8$) relies on the six layers of high resolution silicon detectors of the Inner Tracking System (ITS), with the two innermost layers equipped with Silicon Pixel Detecetors (SPD), a large volume Time Projection Chamber (TPC), and a high granularity Transition-Radiation Detector (TRD). Thanks to its vertex reconstruction, tracking and particle identification capabilities, ALICE can perform a direct charm $v_2$  measurement through fully reconstructed D meson hadronic decays. In these proceedings we present the result on the measurement of D meson $v_2$ from $\sim9.5\cdot10^6$ ($\sim7.1\cdot10^6$) Pb-Pb collisions in the centrality class 30-50\% (15-30\%). 

\section{Event selection and event plane determination}
The data sample used for the D meson $v_2$ analysis has been collected with a minimum bias trigger given by the coincidence of signal in the VZERO scintillators and the SPD, and a semi-central trigger tuned to select events with centrality 0-50\%. 
The centrality selection of the events is done on the distribution of signal in the VZERO scintillators. A Glauber fit is performed on MonteCarlo simulations to extract the total cross section and normalise \cite{alberica}. The event plane is measured from the distribution of charged tracks in the $0<\eta<0.8$ region using the equation 
\begin{equation}
\Psi =\frac{1}{2} \tan^{-1}\left(\frac{\sum_{i=0}^{N} w_i \sin 2\phi_i}{\sum_{i=0}^{N} w_i \cos 2\phi_i}\right)
\end{equation}
where $\Psi$ is the second harmonic event plane and $\phi_i$ is the angle of the $i^\text{th}$ track in the ALICE reference frame \cite{poskEP}. Weights are applied to the tracks depending on their angle in the ALICE reference frame to account for different efficiencies among TPC sectors. To compute the event plane resolution each event is splitted in 2 sub-events with each track in the $0<\eta<0.8$ range randomly assigned to one of the sub-events and the sub-event plane angle $\Psi_{a/b}$ is computed. From the two sub-events plane angles the resolution is computed following the prescriptions in \cite{poskEP}. The resulting event plane is flat to about 2\% level and its resolution is 0.86 in the centrality class 30-50\% and 0.9 in 15-30\% centrality.

\section{D meson reconstruction and $v_2$ extraction}
Open charm hadrons reconstruction at ALICE is based on the invariant mass analysis of fully reconstructed hadronic decay topologies. The D meson $v_2$ has been measured for the decay channels $\text{D}^+\rightarrow K^-\pi^+\pi^+$, $\text{D}^0\rightarrow K^-\pi^+$ and the $\text{D}^{*+}\rightarrow \text{D}^0\pi^+$ analysis is ongoing. As the mean proper decay length c$\tau$ of these particles is of the order of $\sim150$ (D\textsuperscript{0})$-300$ (D\textsuperscript{+})$\mu m$, the strategy to reduce the large combinatorial background is based on the selection of displaced-vertex topologies, i.e. with 
large separation between the primary and the secondary vertices and on the good alignment between the reconstructed D meson momentum and the flight line. 
Particle identification is provided by time of flight measurements in the Time Of Flight (TOF) detector and by specific energy deposit in the TPC. The PID selection strategy was defined with the aim of preserving all the signal. Further analysis details can be found in \cite{Draa}. 
The invariant mass distribution of the reconstructed D meson candidates is splitted in the in-plane and out-of-plane regions defined as the regions $\left[0<\Delta\phi<\frac{\pi}{4}\right)\bigcup \left[\frac{3\pi}{4}<\Delta\phi<\pi\right)$ and $\left[\frac{\pi}{4}<\Delta\phi<\frac{3\pi}{4}\right)$ respectively, 
where $\Delta\phi$ is the azimuthal angle of the reconstructed candidate with respect to the event plane.
An invariant mass analysis is used to extract the amount of signal candidates measured in the two regions ($N_\text{IN}$, $N_\text{OUT}$). A Gaussian distribution is assumed for the signal shape. In each $p_\text{T}$ bin the width of the Gaussian is fixed to the value obtained fitting the $\Delta\phi$ integrated invariant mass distribution. From the values of $N_\text{IN}$ and $N_\text{OUT}$ it is possible to directly compute the elliptic flow as
\begin{equation}
 v_2=\dfrac{\pi}{4}\dfrac{N_{IN} - N_{OUT}}{N_{IN} + N_{OUT}}.
\label{eq:v2}
\end{equation}
Other methods to extract $v_2$ based on 2-particles correlations \cite{qcumulants} have been implemented. Their results have been found in agreement with the $v_2$ obtained from equation \ref{eq:v2}.
\section{Results}
\begin{figure}[htb]
\begin{minipage}[t]{.5\textwidth} 
\centering
\captionsetup{margin={.05\textwidth, 3pt}}
\includegraphics[width=\textwidth]{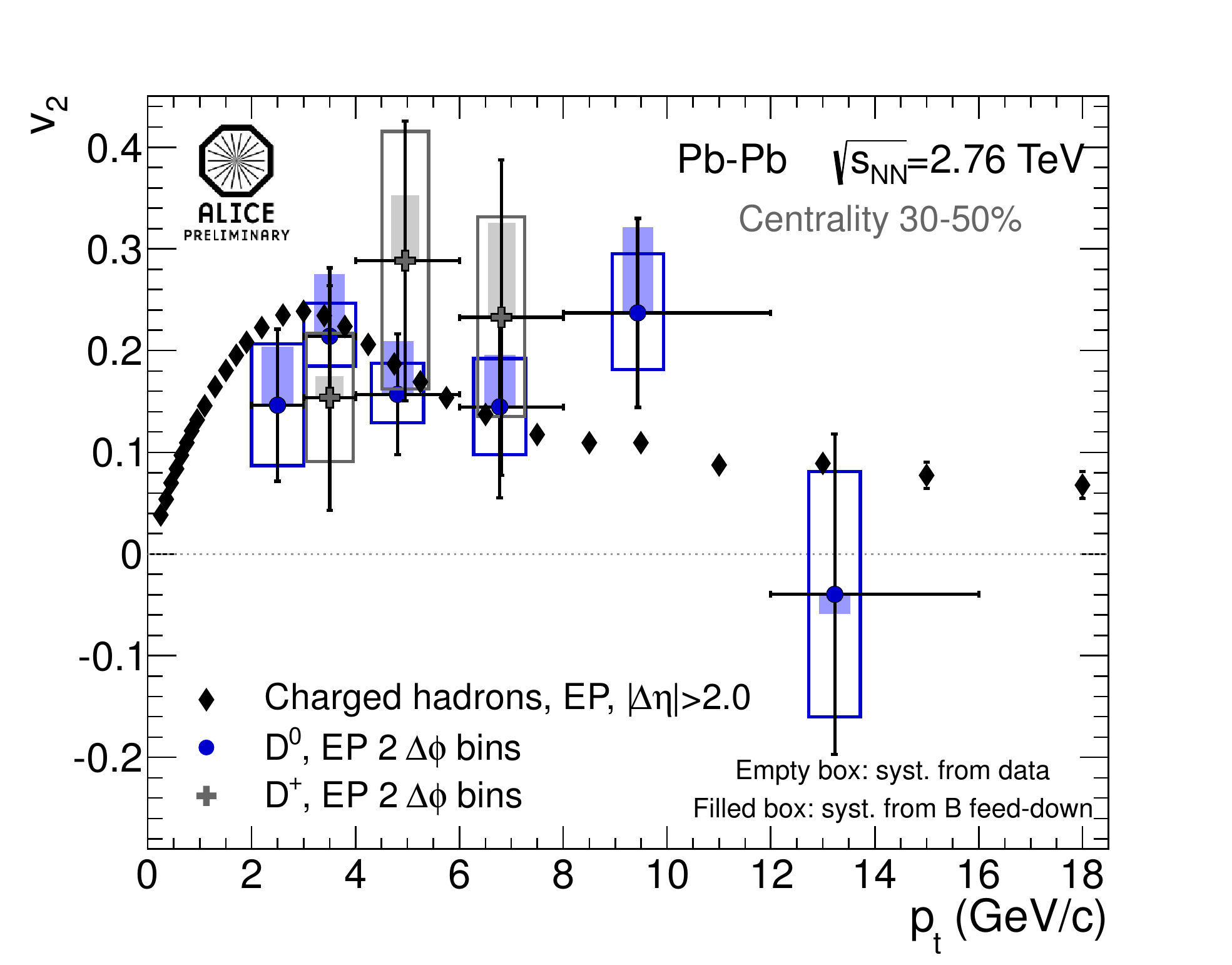}\\
\caption{$\text{D}^0$ (blue circles) and $\text{D}^+$ (grey crosses) $v_2$ measured by ALICE in the 30-50\% centrality class. The empty boxes show the systematic uncertainties from data, the shaded area the uncertainty on the B feed-down. Black points are charged hadron $v_2$.}\label{fig:d0dplus}
\end{minipage}
\begin{minipage}[t]{.5\textwidth} 
\centering
\captionsetup{margin={3pt, .05\textwidth}}
\includegraphics[width=\textwidth]{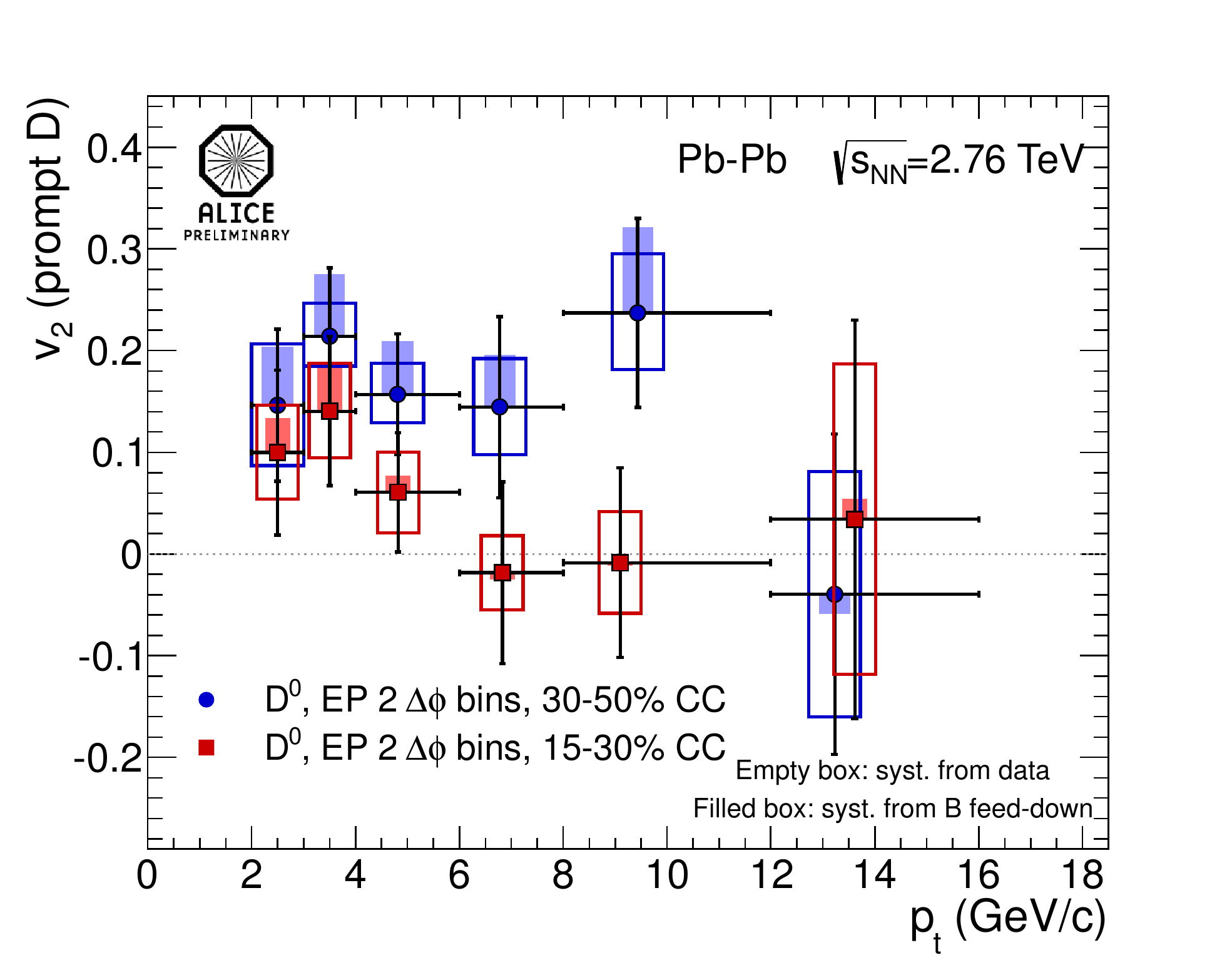}\\
\caption{$\text{D}^0$ $v_2$ measured by ALICE in the 30-50\% (blue circles) and 15-30\% (red squares) centrality classes. The empty boxes show the systematic uncertainties from data, the shaded area the uncertainty on the B feed-down.}\label{fig:centdep}
\end{minipage}
\end{figure}
\subsection{Systematic uncertainties}
Several sources of systematic uncertainties have been considered in the analysis. The main contributions to the total systematic uncertainty come from uncertainties on the yield extraction and from the topological cut selection. The systematic uncertainty on the yield extraction is estimated using different background functions in the fitting procedure and allowing the Gaussian width to vary between in-plane and out-plane invariant mass distributions. The analysis was also repeated using three different sets of topological cuts. Yield extraction and cut variation gave a combined total systematic uncertainty of absolute value 0.02-0.05 each (depending on the channel and on $p_\text{T}$) for the $\text{D}^0$ analysis. Other minor sources of systematic uncertainties are the centrality dependence of the resolution in the centrality range considered, that introduces a systematic uncertainty of $\sim$3\% and a 7\% asymmetric systematic uncertainty towards larger values of $v_2$ that comes from the definition of the sub-events.
\subsection{B feed-down subtraction}
The signal contains a fraction of D meson coming from B decays. Thus the measured elliptic flow $v_2^\text{obs}$ is a combination of the elliptic flows of prompt D ($v_2^\text{prompt}$) and D from B feed-down ($v_2^\text{feed-down}$) mesons. It can be expressed as $v_2^\text{obs}=f_\text{prompt}v_2^\text{prompt}+(1-f_\text{prompt})v_2^\text{feed-down}$, where $f_\text{prompt}$ is the fraction of prompt D meson in the sample. From MonteCarlo simulations and FONLL \cite{FONLL98} predictions we estimated $f_\text{prompt}$ to be in the range 0.7-0.95 depending on channel and $p_\text{T}$, as well as on the topological cuts applied. The relative $R_{\text{AA}}(p_{\text{t}})=\frac{1}{<T_{\text{AA}}>}\frac{dN_{\text{AA}}/dp_{\text{t}}}{d\sigma_{pp}/dp_{\text{t}}}$ suppression of feed-down and prompt candidates also affects the value of $f_\text{prompt}$. The effect was estimated by computing the prompt fraction varying this ratio in the range $0.5<R_{AA}^\text{prompt}/R_{AA}^\text{feed-down}<2$. Most of the 
models predict $0\leq v_2^\text{feed-down}\leq v_2^\text{prompt}$ due to the larger mass of the b quark. For our measurement we make the assumption that $v_2^\text{feed-down}=v_2^\text{prompt}$. This assumption introduces an asymmetric systematic uncertainty on our measurement towards higher $v_2$ values. The maximum value of this uncertainty is given by $v_2^\text{prompt}/f_\text{prompt}$, that corresponds to the limit in which $v_2^\text{feed-down}=0$.
\subsection{D meson elliptic flow}
\begin{figure}[htb]
\begin{minipage}[b]{.5\textwidth} 
\centering
\captionsetup{margin={.05\textwidth, 3pt}}
\includegraphics[width=\textwidth]{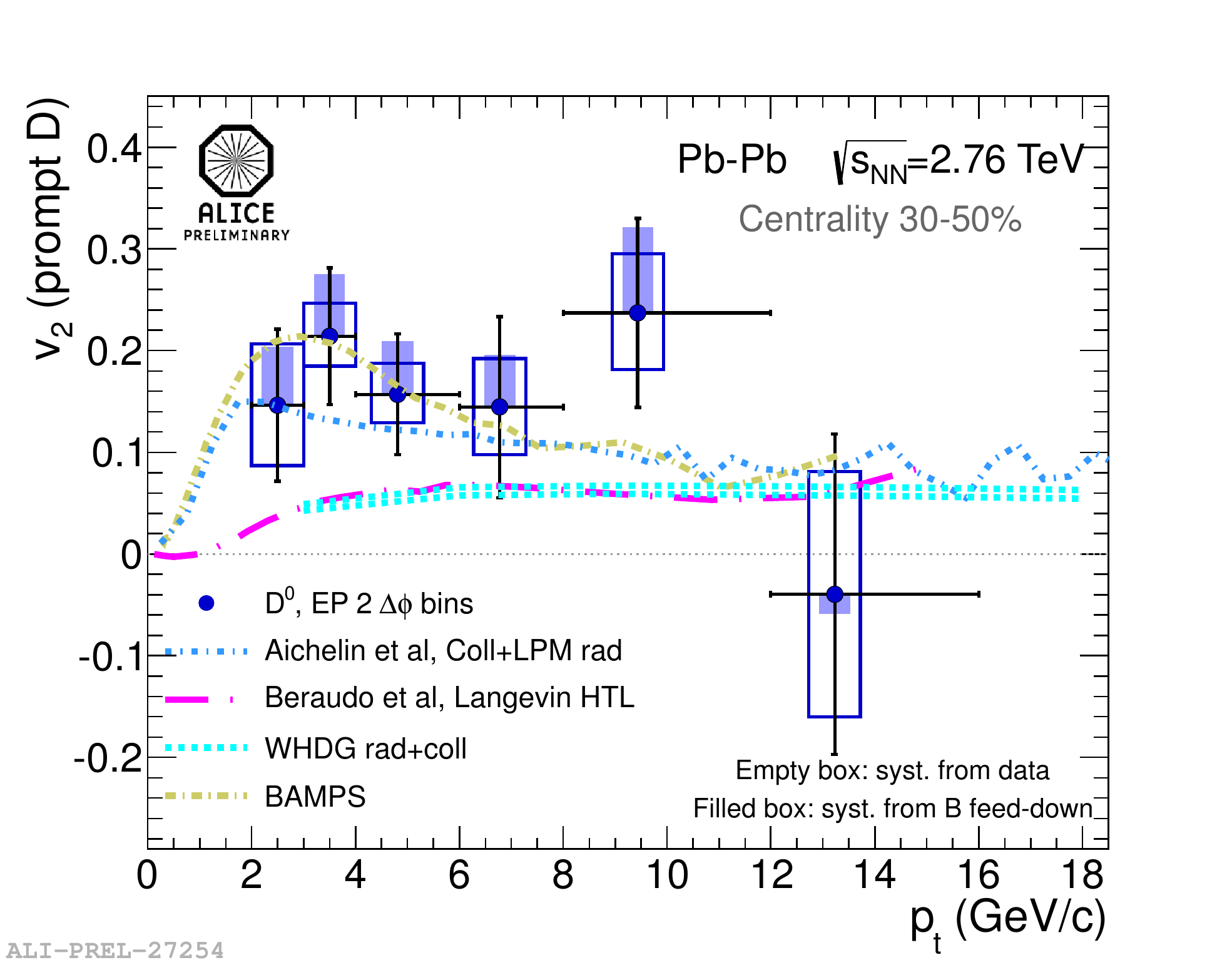}\\
\end{minipage}
\begin{minipage}[b]{.5\textwidth} 
\centering
\captionsetup{margin={3pt, .05\textwidth}}
\includegraphics[width=.9\textwidth]{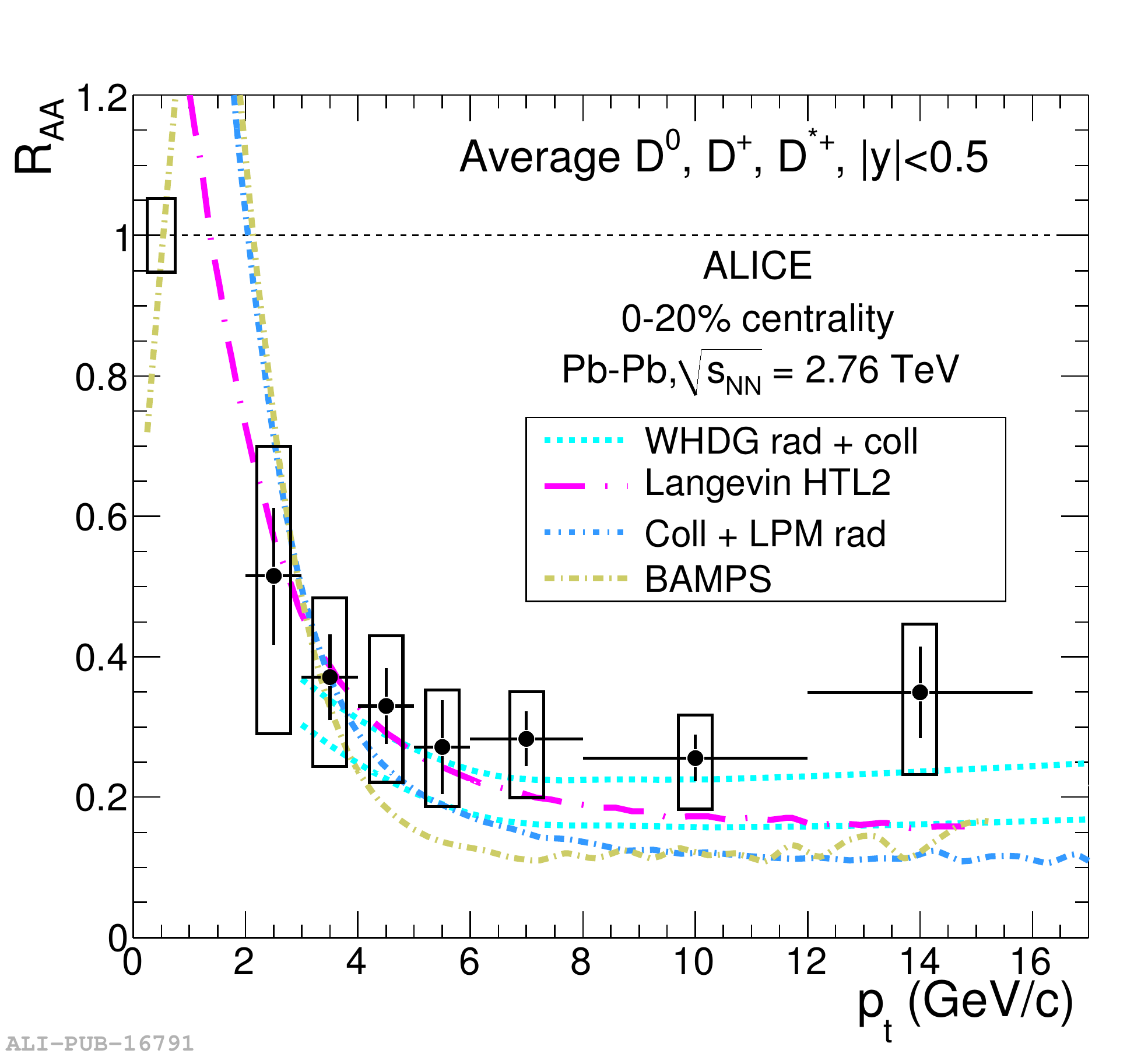}\\
\end{minipage}
\caption{Left: $\text{D}^0$ $v_2$ measured by ALICE in the 30-50\% centrality class (blue circles). The empty boxes show the systematic uncertainties from data, the shaded area the uncertainty on the B feed-down contribution. Right: D meson $R_{AA}$ measurement \cite{Draa} (black). The two measurements are compared to theoretical models \cite{aichelin,beraudo,whdg,bamps}.}\label{fig:models}
\end{figure}

The D meson elliptic flow was measured at ALICE for the $\text{D}^0\rightarrow K^-\pi^+$ and $\text{D}^+\rightarrow K^-\pi^+\pi^+$ channels in the centrality class 30-50\%. The results are shown in figure \ref{fig:d0dplus} for several $p_\text{T}$ bins in the range $2<p_\text{T}<16$~GeV/$c$. In the range where also $\text{D}^+$ measurement is available ($3<p_\text{T}<8$~GeV/$c$), the two measurements are compatible within statistical uncertainties and both are compatible with the ALICE measurement of charged hadron $v_2$ \cite{alicev2} in the same rapidity region. The $\text{D}^0$ $v_2$ measurement was also carried out in the centrality class 15-30\% for $2<p_\text{T}<16$~GeV/$c$. The result is shown in figure \ref{fig:centdep} compared to the result in 30-50\%. Our measurement has been compared to theoretical predictions \cite{aichelin,beraudo,whdg,bamps}, shown in figure \ref{fig:models} (left). The same models have been compared with the D meson $R_{AA}$ measurement from ALICE (figure \ref{fig:models}, 
right).

\section{Conclusions}
In this paper we presented a measurement of the D meson $v_2$ in Pb-Pb collisions at $\sqrt{s_\text{NN}} = 2.76$~TeV collected by the ALICE experiment. The results indicate a non-zero $v_2$ with 3$\sigma$ significance in the $p_\text{T}$ range $2<p_\text{T}<6$~GeV/$c$. The charm $v_2$ is also compatible, within uncertainties, with the $v_2$ of charged hadrons in the same $p_\text{T}$ and rapidity region. 
The data collected in the centrality classes 15-30\% and 30-50\% suggest a decrease of the charm $v_2$ with the collision centrality.
Models of heavy quark transport in the medium can describe the data. However, a simultaneous description of D meson $v_2$ and $R_{AA}$ is still lacking.




\bibliographystyle{elsarticle-num}
\bibliography{bibproceedings}

\begin{thebibliography}{10}
\expandafter\ifx\csname url\endcsname\relax
  \def\url#1{\texttt{#1}}\fi
\expandafter\ifx\csname urlprefix\endcsname\relax\def\urlprefix{URL }\fi
\expandafter\ifx\csname href\endcsname\relax
  \def\href#1#2{#2} \def\path#1{#1}\fi

\bibitem{alice}
K.~Aamodt, et~al., {The ALICE experiment at the CERN LHC}, JINST 3 (2008)
  S08002.

\bibitem{v2}
J.-Y. Ollitrault, {Relativistic hydrodynamics for heavy-ion collisions},
  Eur.J.Phys. 29 (2008) 275--302.

\bibitem{rhicflowRapp}
H.~van Hees, V.~Greco, R.~Rapp, {Heavy-quark probes of the quark-gluon plasma
  at RHIC}, Phys.Rev. C73 (2006) 034913.

\bibitem{Molnar:2004ph}
D.~Molnar, {Charm elliptic flow from quark coalescence dynamics}, J.Phys.G G31
  (2005) S421--S428.

\bibitem{phenixcraa}
A.~Adare, et~al., {Energy Loss and Flow of Heavy Quarks in Au+Au Collisions at
  $s_{\rm NN}^{1/2}$ = 200-GeV}, Phys.Rev.Lett. 98 (2007) 172301.

\bibitem{alberica}
A.~Toia, {Bulk Properties of Pb-Pb collisions at $\sqrt{s_{\rm NN}}$ = 2.76 TeV
  measured by ALICE}, J.Phys.G G38 (2011) 124007.
\newblock \href {http://arxiv.org/abs/1107.1973} {\path{arXiv:1107.1973}}.

\bibitem{poskEP}
A.~M. Poskanzer, S.~Voloshin, {Methods for analyzing anisotropic flow in
  relativistic nuclear collisions}, Phys.Rev. C58 (1998) 1671--1678.

\bibitem{Draa}
B.~Abelev, et~al., {Suppression of high transverse momentum D mesons in central
  Pb-Pb collisions at $\sqrt{s_{NN}}=2.76$ TeV}\href
  {http://arxiv.org/abs/1203.2160} {\path{arXiv:1203.2160}}.

\bibitem{qcumulants}
A.~Bilandzic, R.~Snellings, S.~Voloshin, {Flow analysis with cumulants: Direct
  calculations}, Phys.Rev. C83 (2011) 044913.
\newblock \href {http://arxiv.org/abs/1010.0233} {\path{arXiv:1010.0233}}.

\bibitem{FONLL98}
M.~Cacciari, M.~Greco, P.~Nason, {The P(T) spectrum in heavy flavor
  hadroproduction}, JHEP 9805 (1998) 007.

\bibitem{alicev2}
K.~Aamodt, et~al., {Elliptic flow of charged particles in Pb-Pb collisions at
  2.76 TeV}, Phys.Rev.Lett. 105 (2010) 252302.

\bibitem{aichelin}
P.~Gossiaux, J.~Aichelin, T.~Gousset, V.~Guiho, {Competition of Heavy Quark
  Radiative and Collisional Energy Loss in Deconfined Matter}, J.Phys.G G37
  (2010) 094019.
\newblock \href {http://arxiv.org/abs/1001.4166} {\path{arXiv:1001.4166}}.

\bibitem{beraudo}
W.~Alberico, A.~Beraudo, A.~De~Pace, A.~Molinari, M.~Monteno, et~al.,
  {Heavy-flavour spectra in high energy nucleus-nucleus collisions},
  Eur.Phys.J. C71 (2011) 1666.
\newblock \href {http://arxiv.org/abs/1101.6008} {\path{arXiv:1101.6008}}.

\bibitem{whdg}
W.~Horowitz, {Testing pQCD and AdS/CFT Energy Loss at RHIC and LHC}, AIP
  Conf.Proc. 1441 (2012) 889--891.
\newblock \href {http://arxiv.org/abs/1108.5876} {\path{arXiv:1108.5876}}.

\bibitem{bamps}
O.~Fochler, J.~Uphoff, Z.~Xu, C.~Greiner, {Jet quenching and elliptic flow at
  RHIC and LHC within a pQCD-based partonic transport model}, J.Phys.G G38
  (2011) 124152.
\newblock \href {http://arxiv.org/abs/1107.0130} {\path{arXiv:1107.0130}}.

\end{thebibliography}







\end{document}